\documentclass[12pt]{article}

\usepackage{scicite}
\usepackage{graphicx}

\usepackage{times}
\usepackage{float}
\usepackage{color}

\newenvironment{sciabstract}{%
\begin{quote} \bf}
{\end{quote}}

\newcounter{lastnote}

\title{3D Computational Ghost Imaging} 

\author
{B. Sun,$^{1\ast}$ M. P. Edgar,$^{1}$ R. Bowman,$^{1}$ L. E. Vittert,$^{2}$ S. Welsh,$^{1}$\\ A. Bowman,$^{2}$ and M. J. Padgett$^{1}$\\
\\
\normalsize{$^{1}$SUPA, School of Physics and Astronomy, University of Glasgow, Glasgow, G12 8QQ,}\\
\normalsize{$^{2}$School of Mathematics and Statistics, University of Glasgow, Glasgow, G12 8QQ}\\
\\
\normalsize{$^\ast$To whom correspondence should be addressed: E-mail: b.sun.1@research.gla.ac.uk.}
}

\date{}

\begin{document} 

\baselineskip24pt

\maketitle 

\begin{sciabstract}
Computational ghost imaging retrieves the spatial information of a scene using a single-pixel detector.  By projecting a series of known random patterns and measuring the back-reflected intensity for each one, it is possible to reconstruct a 2D image of the scene.  In this work we overcome previous limitations of computational ghost imaging and capture the 3D spatial form of an object by using several single-pixel detectors in different locations.  From each detector we derive a 2D image of the object that appears to be illuminated from a different direction, using only a single digital projector as illumination.  Comparing the shading of the images allows the surface gradient and hence the 3D form of the object to be reconstructed. We compare our result to that obtained from a stereo-photogrammetric system utilizing multiple high-resolution cameras.  Our lowÐcost approach is compatible with consumer applications and can readily be extended to non-visible wavebands.

\end{sciabstract}

\newpage
Computational ghost imaging (GI) is an alternative technique to conventional imaging and removes the need for a spatially resolving detector. Instead, ghost imaging infers the scene by correlating the known spatial information of a changing incident light field with the reflected (or transmitted) intensity.

The principles of GI were originally demonstrated using spatially entangled photon pairs produced by spontaneous parametric down-conversion, known as quantum GI \cite{Pittman1995PRA,Strekalov1995PRL}.  The two photons propagate along different paths: in the first path, the photon interacts with the object and if not absorbed is detected by a detector with no spatial resolution, in the second path the photon never interacts with the object, but its transverse position is measured by a scanning imaging system. It is by correlating the coincidence measurements over many photon pairs that enables an image of the object to be deduced.

It was subsequently demonstrated that GI could be performed not only using an entangled light source but also with thermal light, a technique commonly termed classical GI \cite{Bennink2002PRL,Gatti2004PRL,Valencia2005PRL,Ferri2005PRL}). In classical GI a copy of the light field is usually made with a beam splitter, one copy of the light field interacts with the object and a non spatially resolving detector and the other copy is recorded with a camera. Correlations between the two detectors again yield an image, albeit one with a higher background than in the quantum case \cite{Jack2009PRL}. The earlier controversy over the distinction between classical and quantum ghost imaging is now largely resolved \cite{ShapiroBoyd2012}.

Classical GI systems can be simplified by introducing a device capable of generating computer programmable light fields, which negates the requirement for the beam splitter and the camera Ð since knowledge of the light field is held in the computer memory. This type of system, termed computational GI \cite{Shapiro2008PRA,Bromberg2009PRA}, has previously been performed using a programmable spatial light modulator (SLM) and a laser, but can also be achieved using a presentation-type programmable light projector. We note that in this form ghost imaging is related to the field of single-pixel cameras \cite{SPI}, the difference being an interchange of the light source and detector.

In both single pixel cameras and ghost imaging systems, inverting the known patterns and the measured intensities is a significant computational problem. A number of sophisticated algorithms have been developed over the years to improve the signal-to-noise ratio (SNR) for different GI systems \cite{Katz2009APL,Ferri2010PRL} but with appropriate normalization \cite{Sun2012OE} a simple iterative algorithm was adopted for this experiment.

All previous GI experiments have been restricted to relatively small 2-dimensional (2D) images, mainly of 2D template objects or 2D outlines of 3-dimensional (3D) objects. In this work we overcome previous limitations of computational GI and capture the 3D spatial form of an object by using several single-pixel detectors in different locations. A 2D image is derived from each detector but appears as if it is illuminated differently. Comparing the shading information in the images allows the surface gradient and hence the 3D form of the surface to be reconstructed.

\begin{figure}[tb]
\centering\includegraphics[width=15cm]{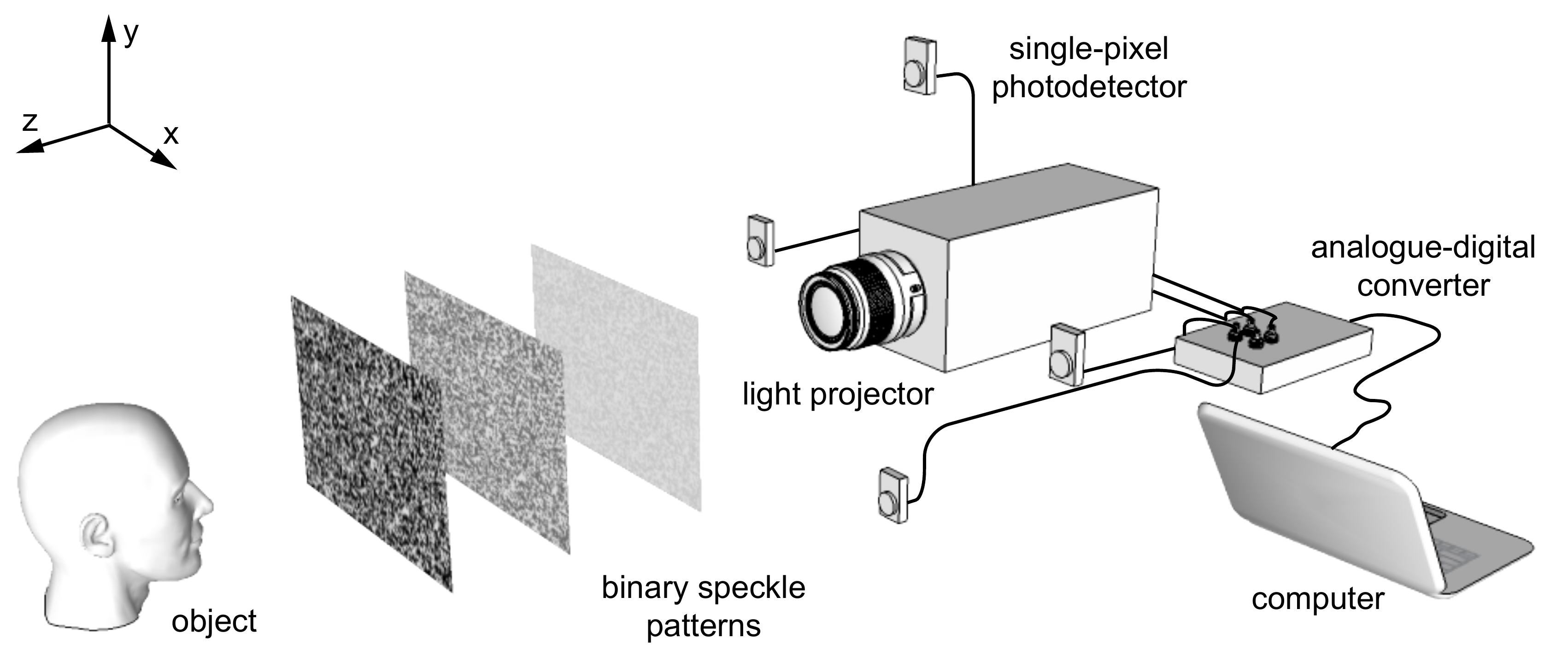}
\caption{Illustration of the experimental setup used for 3D surface reconstructions. The light projector illuminates the object (head) with computer generated random binary speckle patterns. The light reflected from the object is collected on 4 spatially separated single pixel photodetectors. The signals from the photodetectors are measured by the computer via the analogue-digital converter, and used to reconstruct a ghost image for each photodetector.}
\label{Figure1}
\end{figure}

\vspace{1cm}
The experimental setup, illustrated in Fig.\ref{Figure1}, consists of a digital light projector (DLP) (Texas Instruments DLP LightCommander) to illuminate objects with random binary light patterns, four spatially separated single-pixel photodetectors to measure the intensity of the reflected light, an analogue-to-digital converter to digitize the photodetector signals and a computer to generate the random speckle pattern as well as perform 3D reconstructions of the test object.

The digital light projector comprises of a red, green and blue light emitting diode illumination source and a digital micro-mirror device (DMD) to generate the structured illumination. The potentially large operational bandwidth of the DMD ($300\,\rm{nm}$-$2\,\mu\rm{nm}$) however would enable the use of this technique at other wavelengths that are potentially unsuitable for existing imaging technologies.

The projected patterns we use are randomly distributed binary patterns, having a black and white ratio $1:1$, which are projected onto the object using a Nikon 50mm focal length lens. The life-sized mannequin head is positioned about 1m from the lens so that it fits within the projected pattern. Four spatially separated single-pixel photodetectors are positioned in the plane of the lens, separated by 500mm, to record the back-scattered light.  For every binary pattern projected, the corresponding object intensity is measured by each photodetector, which is fed to a computer algorithm.

DMD-based projectors create colour images by displaying 24 binary images (Òbit planesÓ) per frame in quick succession.  By alternating between a binary pattern and its inverse in subsequent bit planes we can demodulate the measured signal at the frequency of the bit plane projection (1440Hz) to isolate the back-reflected signal from light sources at other frequencies such as room lighting. Importantly, the fact that the speckle pattern has equal numbers of black and white pixels enables normalization of the measured signals for each pattern, which has been shown to improve the SNR of the final reconstruction \cite{Sun2012OE}.

\vspace{1cm}
In all iterative GI techniques a 2D representation of the object is reconstructed by averaging the product between the measured photodetector signal and the incident speckle pattern over many patterns. A sequence of $N$ binary patterns, $P_{i}(x,y)$ are reflected from the object, giving a sequence of measured signals $S_{i}$. The 2D reconstruction, $I(x,y)$, which provides an estimate of the object, can be found as 
\begin{equation}
\label{GI}
I(x,y) = \left<\left(S_i-\left<S_i\right>\right)\left(P_i(x,y)-\left<P_i(x,y)\right>\right)\right>,
\end{equation}
where $<.> \equiv \frac{1}{N}\Sigma_i$ denotes an ensemble average for $N$ iterations.

\begin{figure}[H]
\centering\includegraphics[width=13cm]{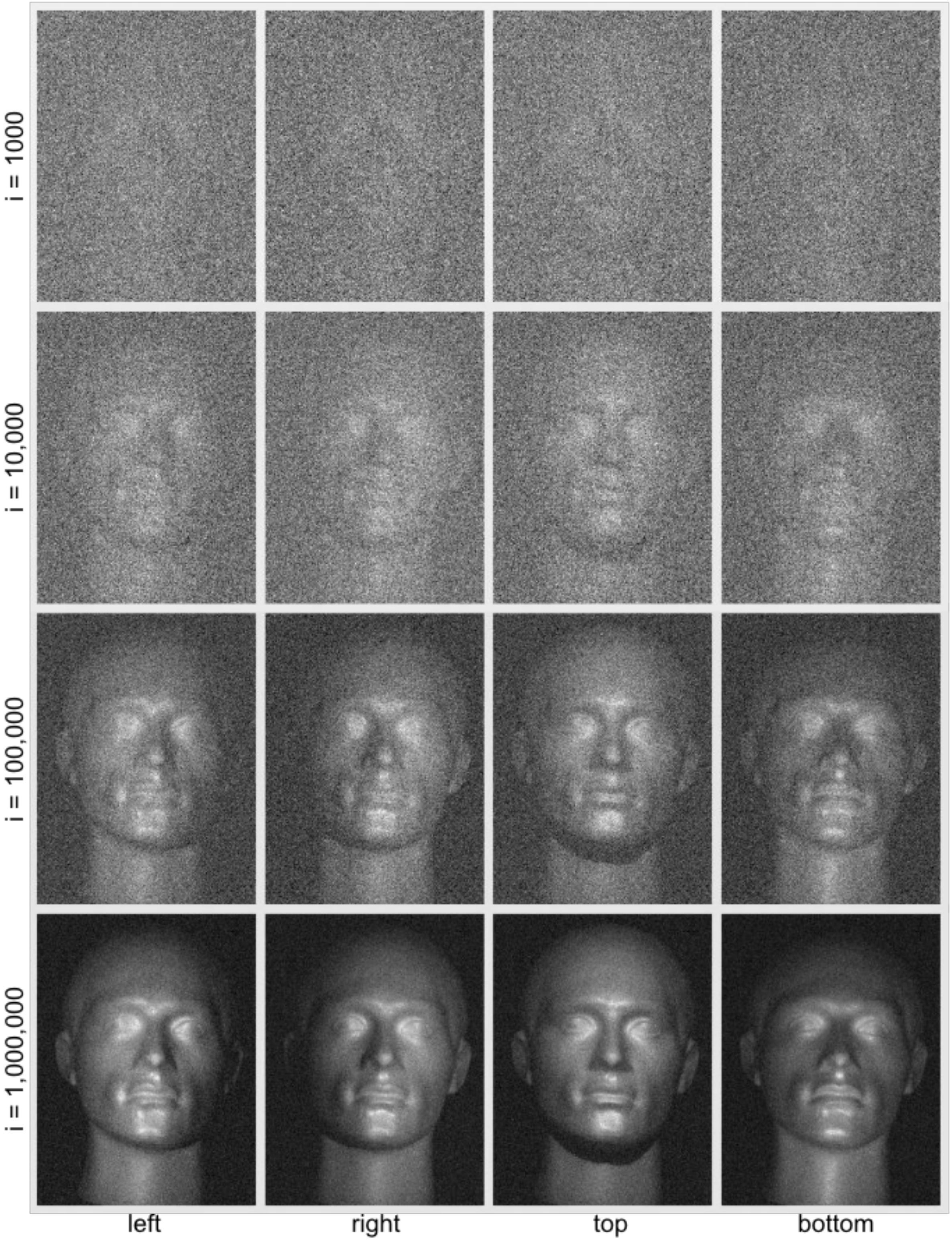}
\caption{Source ghost images from each photodetector employed in the system, reconstructed using the TGI iterative algorithm iteratively from 1000 to 1,000,000. The spatial information in each image is identical, however the apparent illumination source is determined by the location of the relevant photodetector, indicated underneath.}
\label{Figure2}
\end{figure}

Using Eq. \ref{GI} we obtain 2D reconstructions of the object for each of the four photodetectors employed in our system, shown in Fig.\,\ref{Figure2}. The spatial information in each image is identical, however the apparent lighting of the scene is dependent on the location of the detector used to record the back-scattered light (optical imaging systems are reciprocal). Thus, in contrast to imaging systems based on multiple cameras, the perspective of the single pixel detector does not render geometrical distortion to the object being imaged. Instead the location of the detectors introduces the effective lighting position for the scene.

\vspace{1cm}
Depth information of a scene is otherwise lost in a 2D image, but there are instances where it can be inferred  using a technique called `shape from shading' (SFS) \cite{Horn1977, Zhang1999}. From a single image, with one source of illumination, this method relies on the shading caused by geometrical features to reveal the depth of the scene. In general, most SFS methods assume the object exhibits uniform Lambertian (matte) reflectance and that a single light source is located at infinity, such that the incoming lighting vector is constant across the surface of the object. An extension of this technique, called `photometric stereo' (PS) \cite{Woodham1980}, adopts the same assumptions but uses multiple images, each with a different illumination, and taken from the same view point, similar to the types of images retrieved by our 3D computational GI system.

The intensity of a pixel, $I(x,y)$, in the image obtained from the $n^{th}$ detector can be expressed as
\begin{equation}
I_n(x,y) = \alpha\left({\bf L}_n\cdot{\bf N}\right),
\label{IntensityPixel1}
\end{equation}
where $\alpha$ represents the surface reflectance, ${\bf L}_n$ is the unit illumination vector pointing from the detector to the object and ${\bf N}$ is the surface normal unit vector of the object. Thus for all images we can write Eq. \ref{IntensityPixel1} in matrix notation as
\begin{equation}
{\bf I} = \alpha\cdot{\bf L}\cdot{\bf N},
\label{IntensityPixel2}
\end{equation}

 For any pixel $(x,y)$ the unit surface normal is given
\begin{equation}
{\bf N} = \left(1/\alpha\right){\bf L}^{-1}{\bf I},
\label{SurfaceNormal}
\end{equation}
and the reflectance is given by
\begin{equation}
\alpha = |{\bf L}^{-1}{\bf I}|.
\label{Reflectance}
\end{equation}

From the surface normals calculated at each pixel it is possible to determine the gradient between adjacent pixels from which we obtain the surface geometry by integration. In fact, as we record four images, the problem becomes overconstrained as the surface normals represent only two degrees of freedom per pixel.  We can thus remove our assumption of uniform reflectivity and recover an estimate of the surface reflectance $\alpha$ at the same time as finding the object's shape.

\vspace{1cm}
In our algorithm the surface gradients $dz/dx$ and $dz/dy$ were calculated from the surface normals for each pixel. Starting in the centre of the object and working outwards, the gradients were subsequently integrated to recover a depth-map of the object. The height for each pixel was estimated based on the height and the gradient at each of its nearest-neighbor pixels. Thus, the integration was performed iteratively, each pass over the object estimating the height of all the pixels having at least one nearest neighbor with a height estimate to work from. For pixels that have more than one nearest neighbor with an estimated height, the height is calculated from the average of those pixels. This simple algorithm works from the middle of the object outwards, and is capable of integrating around holes in the object where there is no information.

\begin{figure}[H]
\centering\includegraphics[width = 15cm]{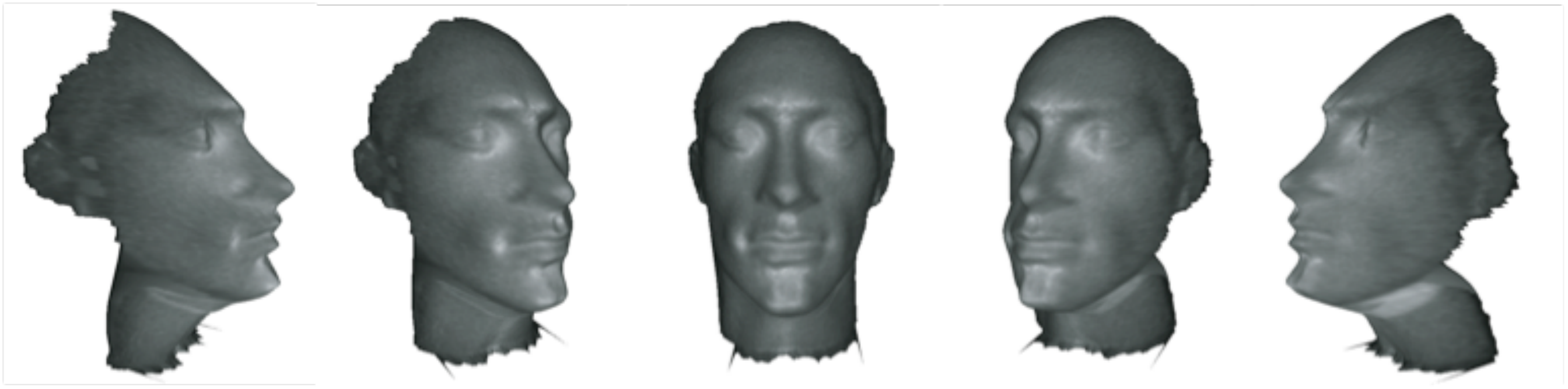}
\caption{Rendered views of the reconstructed facial surface derived by integration of the surface normal data and overlaid with the reflectance data (movie included in supplementary material).}
\label{Figure3}
\end{figure}

Once the algorithm had been appropriately calibrated by imaging a flat surface, accounting for changes of the lighting vector for different pixels across the object plane, the system was tested for objects with geometric complexity. One object investigated was a life-size white polystyrene mannequin head, with approximate dimensions $190\times160\times250\,\rm{mm}$. Using the 2D images shown in Fig.\,\ref{Figure2} we  calibrated for each pixel the reflectance, the surface normals, the surface gradient and the estimated depth as prescribed by our model. A standard 3D graphic package was then used to visualize this profile, overlaid with the reflectance data, as illustrated in Fig.\,\ref{Figure3}. 

To quantify the accuracy of our approach the 3D reconstruction of the test object was compared with a 3D image captured from a stereo-photogrammetric camera system (Di3D$\textsuperscript{\textregistered}$). This latter system uses a matching algorithm on the 2D images from multiple cameras to recover the distance map of an object from the cameras.  The accuracy of this system with facial shapes is well documented \cite{khambay-2008-bjoms} to have a root mean square (RMS) error of order 1mm. 

\begin{figure} [H]
\centering\includegraphics[width = 15cm]{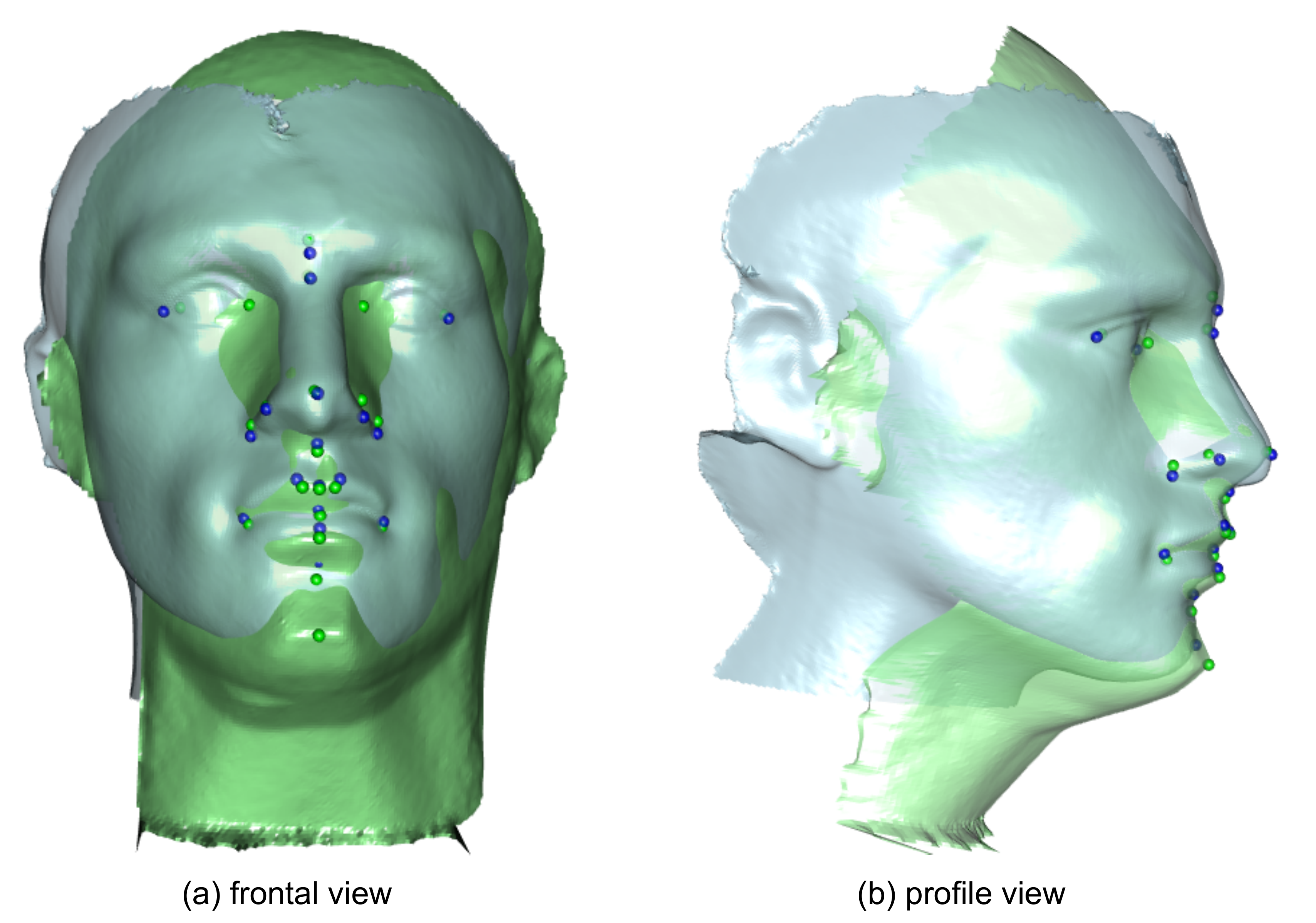}
\caption{The matched ghost imaging (green) and stereo-photogrammetric (blue) reconstructions of the mannequin head, from frontal (a) and profile (b) viewpoints, and with anatomical landmarks (colour coded green and blue respectively) added.}
\label{Figure4}
\end{figure}

To compare the facial profiles measured by the two systems, the shapes are characterised by anatomical landmarks, which are well-defined facial locations \cite{farkas-1994-book} (for example, nose tip, mouth corners etc.). Fig.\ref{Figure4} shows two sets of 21 landmarks superimposed on the facial images by a trained observer. After lateral and angular registration and subsequent depth scaling, the RMS error of our ghost profiler is found to be slightly below 4mm.

Beyond showing that high-quality images of real life objects can be captured using a single-pixel photodetector, our experiment demonstrates that by using a small number of single-pixel detectors, computational ghost imaging methods can give 3D images. 

We have applied this 3D ghost imaging technique to record a facial shape, indicating good quantitative agreement with existing imaging technology based on stereo-photogrammetric systems that employ several high-resolution cameras. An important difference in our approach is that a single projector determines the spatial resolution of the system, removing issues of pixel alignment associated with multiple cameras.  Furthermore, reversing the fundamental imaging process allows for simpler, less expensive detectors to be utilised.   The operational bandwidth of the system is limited not by the efficiency of a pixelated imaging detector but instead by the reflectivity of DMD used for light projection, whose efficiency extends well beyond the visible spectrum.  Development of such technology, for example the use of a broadband white light source, could enable GI systems to become a cheaper alternative for applications in 3D and multi-spectral imaging.

\vspace{1.5cm}
\textbf{Acknowledgments}: We gratefully acknowledge financial support from the UK Engineering and Physical Sciences Research Council for financial support. M.J.P. thanks the Royal Society and the Wolfson Foundation. B.S. and M.P.E. would like to thank Daniele Giovannini for useful discussions. All authors contributed equally to this work and to the writing of the manuscript.

\end{document}